\documentclass[superscriptaddress,preprint]{revtex4}
\usepackage{mathrsfs}
\usepackage{amssymb}
\usepackage[tbtags]{amsmath}
\usepackage{graphicx}
\usepackage{epsfig,graphicx,times}
\usepackage{color}
\usepackage{subfigure}
\usepackage{setspace}
\usepackage{extarrows}

\begin{document}
\title{Nanoparticle acts as a light source to make atom interferometers}
\author{M. Zhang\footnote{zhangmiao079021@163.com}}
\affiliation{School of Physical Science and Technology, Southwest Jiaotong University, Chengdu 610031, China}
\date{\today}

\begin{abstract}
Atomic Kapitza-Dirac (KD) scattering in the classical standing wave of lights is widely used to make the laser-pulsed atom interferometers. In this theoretical work, we show that the dielectric nanoparticle can be used as a weak light source to generate atomic KD scattering and make atom interferometer. We consider a cavity system consisting of atoms and a single nanoparticle, where the nanoparticle is illuminated by the external laser pulses. Atoms and the cavity mirrors are not illuminated by the laser beam, so nanoparticle acts as a pulsed source to excite the cavity mode which non-resonantly excites atomic internal state and generates a weak KD scattering of atomic external motion. We use twice such
scattering to split atomic path and consequently use two classical laser pulses to recombine the separated paths for generating the Ramsey-Bord\'{e} atom interferometer.
This atom interferometer is entangled with the centre-of-mass motion of nanoparticle and could be developed to search the small nonclassicality of motional nanoparticle, because the cavity mode excitation is associated with nanoparticle's position relative to the node or antinode of that optical mode.
\end{abstract}

\maketitle
\section{Introduction}
The nanometer-size dielectric particle (nanoparticle) consisting of millions of atoms has been proposed to test quantum gravity and the spontaneous collapse models of wave function~\cite{Romero-Isart-PRL,Romero-Isart-collapse,
Quantum gravity1,Quantum gravity2,RMP,NP-decoherence,NP-Gdecoherence,CQG-decoherence}.
The advantage on quantum gravity testing can be easily seen from the Newtonian gravitational potential energy, $Gm_1m_2/r$, where $r$ is the distance between two massive spheres $m_1$ and $m_2$, and $G=6.67\times10^{-11} {\rm N}\cdot{\rm m}^2\cdot{\rm kg}^{-2}$ the gravitational constant. This energy is equivalent to that of a microwave photon $\hbar\omega_m$ of frequency $\omega_m=Gm_1m_2/(\hbar r)$. Using atoms as the test particles, we have $\omega_m=6.4\times10^{-22}$ Hz, with the mass $m_1=m_2=10^{-25}$ kg and the distance $r=10\,\,\mu {\rm m}$ between atoms. It should be extremely difficult to measure such a microwave photon. Note that, the test particles can not be too close for avoiding the electromagnetic interactions, i.e., the Casimir-Polder interaction between neutral particles. Considering if two molecules of mass $m_1=m_2=1.6\times10^{-23}$~kg~\cite{molecular}, the frequency $\omega_m=1.6\times10^{-17}$ Hz is still small. However, if the test particles are two nanoparticles of more mass $m_1=m_2=10^{-15}$~kg, the frequency $\omega_m=0.064$~Hz, or say the phase shift $\omega_mt$ with an effective interaction time $t$, seems to be experimental detectable by using the matter wave interferometry of nanoparticles~\cite{Quantum gravity1,Quantum gravity2}. Basically, the gravitational interaction energy is proportional to the quadratic mass~\cite{Diosi,Penrose,arXivGD}, so the large particles seem to be essential for demonstrating the gravity-induced entanglement~\cite{Quantum gravity1,Quantum gravity2} and the gravitational decoherence~\cite{RMP,NP-decoherence,NP-Gdecoherence,
CQG-decoherence}.

The matter wave interferometers of nanoparticles have been not realized yet, but great progress has been made both in theories and experiments~\cite{Barker2010,Barker2012,EXP-PNAS2010,
EXP-PRL2012,EXP-NJP2013,EXP-PRL2016,EXP-PRL2019-1,
EXP-PRL2019-2,EXP-Science,EXP-Nature,Rep2020,Radim-Filip,Quantum-Sci-Techol,arXiv}, focused on nanoparticles loading, trapping, cooling, and measurement. The experiment~\cite{EXP-Science} shows that the single nanoparticle can be trapped as a harmonic oscillator by the optical tweezer and can be cooled into the motional quantum ground state by using the cavity QED. In that experiment, the position and momentum uncertainties of the cooled nanoparticle are on the orders of $\Delta_x=\sqrt{\hbar/(2 m\omega)}\approx0.25\,\,{\rm \AA}$ and $\Delta_p=\sqrt{m\hbar\omega/2}=m\times13\,\,\mu{\rm m}\cdot{\rm s}^{-1}$, with the vibrational frequency $\omega=2\pi\times80$ kHz and the $m=10^8$ atomic mass units (amu) of harmonic oscillator. Switching off optical trap instantaneously, the nanoparticle will be released as a free fall~\cite{MAQRO2012,MAQRO2015,NC2021,free-N-KD}. This provides opportunities of making the matter wave interferometers of nanoparticles~\cite{Romero-Isart-PRL,Romero-Isart-collapse}.

A major difficulty of making nanoparticle interferometers is the split-recombination of nanoparticle's wave packet. According to Wigner equation~\cite{Wigner1932,MiaoPRD}, i.e., the Schr\"{o}dinger equation in (position, momentum) phase space, the dynamical nonclassicality of centre-of-mass motion of nanoparticle is generated by the high-order gradient potential that the particle underwent. Laser can provide the considerable high-order nonlinear potential for test particles due to the small wave length of light, and thus has been widely used to make atom interferometers. However, this seems insufficient for coherently splitting and recombining nanoparticle's wave packet. Using the atomic mass $m_a=86.9$ amu~\cite{D2} and the wave number $k=2\pi/\lambda=2\pi/(780\,{\rm nm})$ in the usual atom interferometers, one can easily find that the single photon recoil velocity $\hbar k/m\approx5\,\,{\rm nm}\cdot{\rm s}^{-1}$ of nanoparticle is far smaller than that $\hbar k/m_a\approx5.8\,\,{\rm mm}\cdot{\rm s}^{-1}$ of atom, because the nanoparticle of mass $m=10^8$ amu is overweight. The long time free evolutions of nanoparticle in microgravity laboratory~\cite{MAQRO2012,MAQRO2015,NC2021}, such as $10$~s, may generate the experimentally detectable interference fringes, but shall face the obstacles of decoherence, such as the collision with residual air molecules and the blackbody
radiation~\cite{Romero-Isart-collapse}.
The multi-photon scattering of highly transparent nanoparticle would be useful~\cite{NC2021,free-N-KD}. Similar to the Kapitza-Dirac (KD) scatterings of atom~\cite{RMP-AKD,KD-E,ZHU-ATOM} and molecules~\cite{KD-molecules,KD-molecules2019} in the non-resonant standing lights, the multi-photon KD scattering of dielectric nanoparticle would results in many paths of nanoparticle, and thus needs also the high spatial resolution measurements, for example, by using the tightly focused ultraviolet (UV) laser pulse~\cite{MAQRO2015,MAQRO2012}. Different from the above traditional approaches, Romero-Isart et al.~\cite{Romero-Isart-PRL,Romero-Isart-collapse} propose using the feature of wave collapse to generate nanoparticle interference. Based on the entanglement between nanoparticle and photons in a cavity, they measure the output photons for collapsing nanoparticle's state. 

In this article, we design an atom interferometer to search the small nonclassicality of centre-of-mass motion of nanoparticle. The atomic matter wave interferometers have advantages not only on the measurements of gravitational acceleration~\cite{ChuG} but also on the detections of some weak interactions between atom and electromagnetic fields, such as the He-McKellar-Wilkens phase~\cite{HMW} and the relativistic Aharonov-Casher effect~\cite{AC} of motional atoms. The atomic matter waves may be also very sensitive to the weak lights radiated by the dielectric nanoparticle~\cite{AN-PRA,ZJ-PRA,NWJ-OC,NWJ-PRA,AN-Paul}. For that, we consider a hybrid cavity system consisting of an atom and a nanoparticle. The input external laser beam is the pulsed one and directed at the nanoparticle, does not illuminate atom and the cavity mirrors. So nanoparticle acts as a pulsed source to excite cavity modes and consequently generate the atomic KD scattering. This scattering is explained by the position dependent ac Stark effect of atom in the non-resonant standing wave. Atomic internal state is not excited, and so the present KD scattering is also effective for the case that many atoms (atomic cloud) are simultaneously in the cavity. This allowed us to make an atom interferometer, for example, the symmetric Ramsey-Bord\'{e} atom interferometer. This atom interferometer has several uses, demonstrating the cavity QED itself, detecting the classical motions of nanoparticle, and searching the small nonclassicality of nanoparticle, because the cavity mode excitation is related to the centre-of-mass motion of nanoparticle.

The article is organized as follows. In Sec.~II, we use the well-known Jaynes-Cummings (JC) model~\cite{JC} to describe atom-cavity interaction and use the widely recognized theories in ref.~\cite{Romero-Isart2011} to describe the laser-nanoparticle-cavity interaction. Considering atomic internal transition is detuning a litter with the input laser, we use several unitary transformations to derive atomic ac Stark effect (called also two-photon Stark effect). The obtained Stark shift is directly coupled with the centre-of-mass motion of nanoparticle and is the crucial one for generating the nanoparticle dependent KD scattering of atoms. Based on the obtianed KD scattering, in Sec.~III, we design a symmetric Ramsey-Bord\'{e} atom interferometer to measure the possible nonclassicality of motional nanoparticle. The time-evolution operators method is used to derive the signal of the considered atom interferometer. Finally, we present our conclusion in Sec. IV.

\section{The weak KD scattering of atom}
Considering an atom and a nanoparticle being inside a cavity, as shown in Fig.~1, where the nanoparticle is illuminated by a classical laser pulse which propagates along the perpendicular direction of cavity axis. Nanoparticle acts a light source to nearly-resonantly excite one of the cavity modes and consequently generate the atomic ac Stark effect when the transition frequency between two atomic levels is detuning a little with that cavity mode. In this section, we use the unitary transformation approach to derive the considered ac Stark effect.
The obtained ac Stark effect depends not only on the atomic position but also the nanoparticle's position, and thus results in the atom-nanoparticle coupled KD scattering. We temporarily consider the single atom and the single nanoparticle inside the cavity. In the later, we will easily find that the single atom theory is also effective to describe the practical case that many atoms are simultaneously inside the interaction cavity.

\begin{figure}[tbp]
\includegraphics[width=10cm]{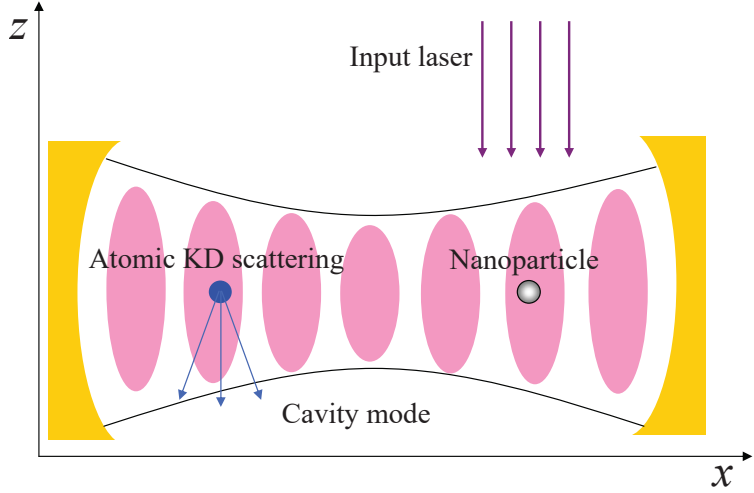}
\caption{Nanoparticle acts as a light source to generate atomic KD scattering. An external classical laser beam (propagating along $z$ direction) is directed at the nanoparticle, does not illuminate atom and the cavity mirrors, so nanoparticle is used as a light source to nearly-resonantly excite one of the cavity modes and generates atomic KD scattering. This scattering is coupled with the centre-of-mass motion of nanoparticle, as the mode excitation is associated with the position of nanoparticle relative to the node or antinode of that mode.}
\end{figure}

\subsection{The Hamiltonian}
The Hamiltonian of the cavity system Fig.~1 is written as
\begin{equation}
\begin{aligned}
\hat{H}=\frac{\textbf{p}_a^2}{2m_a}+\frac{\textbf{p}^2}{2m}+\hat{H}_{\rm AC}
+\hat{H}_{\rm NF}+\hat{H}_{\rm CF}\,.
\end{aligned}
\end{equation}
The $\textbf{p}_a^2/(2m_a)$ and $\textbf{p}^2/(2m)$ are the non-relativistic kinetic energies of atom and nanoparticle, respectively.
The Hamiltonian $\hat{H}_{\rm AC}$ describes the interaction between atom and the cavity mode. The $\hat{H}_{\rm NF}$ describes the interactions between nanoparticle and the electromagnetic fields of the cavity modes, the input classical laser, and the free environment. The $\hat{H}_{\rm CF}$ is the Hamiltonian of cavity modes interacting with the free fields in environment. The interaction between atom and the free fields is negligible, because the atom is not excited throughout the paper.

We use the well-known JC model~\cite{JC} to describe the atom-cavity interaction, i.e.,
\begin{equation}
\begin{aligned}
\hat{H}_{\rm AC}=\hbar\Omega_{a}
(\hat{a}\hat{\sigma}_{+}e^{i\delta_{ac}t}
+\hat{a}^\dagger\hat{\sigma}_{-} e^{-i\delta_{ac}t})\,,
\end{aligned}
\end{equation}
where $\delta_{ac}=\omega_a-\omega_c$ is a small detuning between atom and a selected standing wave between the cavity mirrors. The frequencies of cavity standing waves are on the order of $\omega_n=nc/(2L)\approx n\times7.5$~GHz, with a cavity length $L=2$~cm, for example. The $7.5$~GHz means that, only one cavity mode could be effective to generate the considerable effect of atom, because the detunings between atom and the other cavity modes are too large. This is why we say the selected cavity mode. In the above JC model, the $\hat{a}^\dagger$ and $\hat{a}$ are respectively the bosonic creation and annihilation operators of that cavity mode. The $\hat{\sigma}_+=|e\rangle\langle g|$ and $\hat{\sigma}_-=|e\rangle\langle g|$ are called usually raising and lowering operators, which generate the transition between atomic internal  states. The resonant Rabi frequency $\Omega_{a}$ of atom-cavity interaction is related to the atomic position $(x_a,y_a,z_a)$, for example, $\Omega_{a}=\Omega_{a0}\cos(kx_a)$ with the plane wave approximation, and where $\Omega_{a0}$ is the maximum Rabi frequency at the antinode of the standing wave. The $\cos(kx_a)$ is the key to generate atomic KD scattering, with the optical $k$.

The Hamiltonian of nanoparticle-lights interaction is given by refs.~\cite{Romero-Isart2011,Romero-Isart2019},
\begin{equation}
\begin{aligned}
\hat{H}_{\rm NF}&=-\frac{\epsilon_c\epsilon_0}{2}\int_{V_N}
\textbf{E}(\textbf{x}+\textbf{x}')\cdot\textbf{E}(\textbf{x}+\textbf{x}'){\rm d}^3\textbf{x}'\approx-\frac{\epsilon_c\epsilon_0 V_N}{2}
\textbf{E}(\textbf{x})\cdot\textbf{E}(\textbf{x})\,,
\end{aligned}
\end{equation}
with the permittivity $\epsilon_0$ of vacuum, and where
$\epsilon_c=3(\epsilon_r-1)/(\epsilon_r+2)$, with $\epsilon_r$ being the relative dielectric constant of nanoparticle.
The $\textbf{x}$ is the center of mass of the dielectric sphere (with volume $V_N$), and $\textbf{x}'$ is the relative coordinate of the volume-element ${\rm d}^3\textbf{x}'$ in the dielectric sphere. Considering the radius $R$ of nanoparticle is far smaller than the wavelength of lights, so the integral has been solved by using the point-particle approximation.
The electric field can be written as
\begin{equation}
\begin{aligned}
\textbf{E}(\textbf{x})=\textbf{E}'(\textbf{x},t)+\textbf{E}_c\cos(kx)
(\hat{a}^\dagger e^{i\omega_ct}+\hat{a}e^{-i\omega_ct})+\textbf{E}_l[e^{-i(\omega_l t-k_lz)}+e^{i(\omega_l t-k_lz)}]\,,
\end{aligned}
\end{equation}
where $\textbf{E}'(\textbf{x},t)$ is used to explain the nanoparticle radiated lights in free space and the possible excitations of some other standing waves (different from $\omega_c$) between the cavity mirrors.
The $\textbf{E}_c=\textbf{e}_c\sqrt{\hbar\omega_c/(2\epsilon_0V_c)}$ is the electric field strength of the cavity mode of wave number $k=\omega_c/c$, and where $V_c$ is the volume of cavity. The $\textbf{e}_c$ is the polarization vector of electric field perpendicular to the cavity axis.
The $\textbf{E}_l$ is the electric field strength of the input classical laser beam, which  propagates along $z$ direction, with wave number $k_l=\omega_l/c$ and frequency $\omega_l$.
The plane waves approximation is used, i.e., considering the positional uncertainties of particles are far smaller than the waists of laser and the cavity, so the electric fields $\textbf{E}_c(y,z)\approx\textbf{E}_c$ and $\textbf{E}_l(x,y)\approx\textbf{E}_l$ are approximately invariant during the short interaction time (of laser pulse).

Inserting electric field (4) into Eq.~(3),
we write the interaction Hamiltonian of nanoparticle as
\begin{equation}
\begin{aligned}
\hat{H}_{\rm NF}&=-\hbar\Omega_c\hat{a}^\dagger\hat{a}
-\hbar(\Omega_l\hat{a}^\dagger e^{i\delta_{cl}t}+\Omega_l^*\hat{a} e^{-i\delta_{cl}t})+\widetilde{H}_{\rm NF}\,,
\end{aligned}
\end{equation}
with the laser-cavity detuning $\delta_{cl}=\omega_c-\omega_l$,
the two-photon Rabi frequencies
\begin{equation}
\begin{aligned}
\Omega_{c}=\frac{\epsilon_c\epsilon_0 V_N\textbf{E}_c\cdot\textbf{E}_c}{2\hbar}\cos^2(kx)\,,
\end{aligned}
\end{equation}
and
\begin{equation}
\begin{aligned}
\Omega_{l}=\frac{\epsilon_c\epsilon_0 V_N\textbf{E}_c\cdot\textbf{E}_le^{ik_lz}}{2\hbar}\cos(kx)\,.
\end{aligned}
\end{equation}
The $\widetilde{H}_{\rm NF}$ contains all the other terms in (3), for example, the interaction between nanoparticle and the other cavity modes. Nevertheless, the excitations about the other cavity modes should be weak because the input laser is large detuning with those cavity modes.
According to Eqs.~(1) and (5), we can write the Hamiltonian of the total system as
\begin{equation}
\begin{aligned}
\hat{H}=&\hat{H}_{\rm NCF}+\frac{\textbf{p}_a^2}{2m_a}+\frac{\textbf{p}^2}{2m}+\hbar\Omega_{a}
(\hat{a}\hat{\sigma}_{+}e^{i\delta_{ac}t}
+\hat{a}^\dagger\hat{\sigma}_{-} e^{-i\delta_{ac}t})\\
&
-\hbar\Omega_c\hat{a}^\dagger\hat{a}
-\hbar(\Omega_l\hat{a}^\dagger e^{i\delta_{cl}t}+\Omega_l^*\hat{a} e^{-i\delta_{cl}t})\,,
\end{aligned}
\end{equation}
with $\hat{H}_{\rm NCF}=\hat{H}_{\rm CF}+\widetilde{H}_{\rm NF}$. The second line in Hamiltonian (8) describes the non-resonant excitation of the desired cavity mode $\omega_c$ which is used to generate the ac Stark effect of atom.

\subsection{The unitary transformations approach for atomic ac stark effect}
In order to clearly show the ac Stark effect of atom, we use several unitary transformations to change Hamiltonian (8).
Firstly, we perform an unitary transformation $\exp(i\delta_{c l}t\hat{a}^\dagger\hat{a})$ to the total system, then Hamiltonian (8) becomes
\begin{equation}
\begin{aligned}
\hat{H}_1=&\hat{H}'_{\rm NCF}+\frac{\textbf{p}_a^2}{2m_a}+\frac{\textbf{p}^2}{2m}+\hbar\Omega_{a}
(\hat{a}\hat{\sigma}_{+}e^{i\delta_{al}t}
+\hat{a}^\dagger\hat{\sigma}_{-} e^{-i\delta_{al}t})+\hat{H}_D\,,
\end{aligned}
\end{equation}
with the driving cavity mode
\begin{equation}
\begin{aligned}
\hat{H}_D=\hbar\delta_{cl}'\hat{a}^\dagger\hat{a}
-\hbar(\Omega_l\hat{a}^\dagger+\Omega_l^*\hat{a})\,,
\end{aligned}
\end{equation}
and the detunings $\delta_{al}=\omega_a-\omega_l$ and $\delta_{cl}'=\delta_{cl}-\Omega_c$. The Hamiltonian $\hat{H}'_{\rm NCF}$, with apostrophe, is that after the above unitary transformation, i.e., $\hat{H}'_{\rm NCF}=\exp(-i\delta_{c l}t\hat{a}^\dagger\hat{a})\hat{H}_{\rm NCF}
\exp(i\delta_{c l}t\hat{a}^\dagger\hat{a})$, and the similar presentations will be used in follows for the more unitary transformations.

Using the second unitary evolution $\hat{D}=\exp(-it\hat{H}_{\rm D}/\hbar)$ of the driving cavity mode, we compute the so-called BCH formula ~\cite{Zassenhaus-2012,Zassenhaus-1967}
\begin{equation}
\begin{aligned}
\hat{D}^\dagger\hat{a}^\dagger\hat{D}&=\hat{a}^\dagger
+\frac{it}{\hbar}[\hat{H}_{D},\hat{a}^\dagger]+\frac{1}{2!}(\frac{it}{\hbar})^2
[\hat{H}_{D},[\hat{H}_{D},\hat{a}^\dagger]]+\cdots\\
&=\hat{a}^\dagger
+it\delta_{cl}'(\hat{a}^\dagger-\frac{\Omega_l^*}{\delta_{cl}'})
+\frac{(it\delta_{cl}')^2}{2!}
(\hat{a}^\dagger-\frac{\Omega_l^*}{\delta_{cl}'})+\cdots
\\
&=\hat{a}^\dagger e^{i\delta_{cl}' t}-\frac{\Omega_l^*}{\delta_{cl}'}(e^{i\delta_{cl}' t}-1)\,,
\end{aligned}
\end{equation}
and which can be checked by $\hat{D}^\dagger\hat{H}_D\hat{D}=\hat{H}_D$ with $(\hat{D}^\dagger\hat{a}^\dagger\hat{D})^\dagger=\hat{D}^\dagger\hat{a}\hat{D}$.
Consequently, Hamiltonian (9) becomes
\begin{equation}
\begin{aligned}
\hat{H}_2
=&\hat{H}''_{\rm NCF}+\frac{\textbf{p}_a^2}{2m_a}+\frac{\textbf{p}'^2}{2m}+\hbar\Omega_{a}
(\hat{a}\hat{\sigma}_{+}e^{i\Delta t}
+\hat{a}^\dagger\hat{\sigma}_{-} e^{-i\Delta t})\\
&-
\hbar\Omega_{a}
\left[\eta\hat{\sigma}_{+}(e^{i\Delta t}-e^{i\delta_{al}t})
+\eta^*\hat{\sigma}_{-} (e^{-i\Delta t}-e^{-i\delta_{al}t})\right]\,,
\end{aligned}
\end{equation}
with
\begin{equation}
\begin{aligned}
\Delta=\omega_a+\Omega_c-\omega_c=\delta_{al}-\delta'_{cl}\,,
\end{aligned}
\end{equation}
and
\begin{equation}
\begin{aligned}
\eta=\frac{\Omega_l}{\delta_{cl}'}
=\frac{\Omega_l}{\omega_c-\omega_l-\Omega_c}\,.
\end{aligned}
\end{equation}

The parameter $\eta$ is controlled by $\Omega_l$ (i.e., the power of input classical laser), so we consider some values in the regime of $|\eta|\geq1$. Furthermore, we assume the large detuning condition
$\Omega_{a}\ll\delta_{al}\ll
\Delta$ is satisfied, which implies $\Omega_l\gg\omega_a-\omega_l=\delta_{al}\gg\Omega_{a}$.
Limited within these conditions, the terms associated with $\exp(\pm i\Delta t)$ in Hamiltonian~(12) are negligible, and thus
\begin{equation}
\begin{aligned}
\hat{H}_2
\approx&\hat{H}''_{\rm NCF}+\frac{\textbf{p}_a^2}{2m_a}+\frac{\textbf{p}'^2}{2m}+
\hbar\Omega_{a}
\left(\eta\hat{\sigma}_{+}e^{i\delta_{al}t}
+\eta^*\hat{\sigma}_{-} e^{-i\delta_{al}t}\right)\,.
\end{aligned}
\end{equation}
The last term describes a driving atom but decouples with the cavity mode.
In order to clearly show the effect of that term, we take the third unitary evolution $\exp(i\delta_{al}t|e\rangle\langle e|)$ to change Hamiltonian (15),
\begin{equation}
\begin{aligned}
\hat{H}_3
=&\hat{H}''_{\rm NCF}+\frac{\textbf{p}_a^2}{2m_a}+\frac{\textbf{p}'^2}{2m}
+\hbar\delta_{al}|e\rangle\langle e|+
\hbar\Omega_{a}
\left(\eta\hat{\sigma}_{+}
+\eta^*\hat{\sigma}_{-}\right)\,.
\end{aligned}
\end{equation}
The last term in Hamiltonian (16) describes the resonant excitation of atom, but is suppressed by the detuning $
\hbar\delta_{al}|e\rangle\langle e|$. Thus, we take the unitary transformation
$\exp(-i\hat{R})$ with a weak resonant operator $\hat{R}=\mu\hat{\sigma}_++\mu^*\hat{\sigma}_-$, and where $\mu=-i\eta\Omega_a/\delta_{al}$ is the small complex number and time independent. This unitary transformation can exactly eliminate the resonant coupling $\hbar\Omega_{a}
\left(\eta\hat{\sigma}_{+}
+\eta^*\hat{\sigma}_{-}\right)$ in Eq.~(16), such that the standard Stark effect of atom appears. The computation is based on the following two-order BCH formulas of atom,
\begin{equation}
\begin{aligned}
e^{i\hat{R}}
\hat{\sigma}_+e^{-i\hat{R}}
&=\hat{\sigma}_+
+i[\hat{R},\hat{\sigma}_+]
+\frac{i^2}{2!}
[\hat{R},
[\hat{R},\hat{\sigma}_+]]+\mathcal{O}
(|\mu|^3)\\
&\approx\hat{\sigma}_+
-i\mu^*\hat{\sigma}_z
+\mu^*\mu^*\hat{\sigma}_--\mu^*\mu\hat{\sigma}_+
\\
&=\hat{\sigma}_+
+\frac{\eta^*\Omega_a}{\delta_{al}}\hat{\sigma}_z
+\mu^*\mu^*\hat{\sigma}_--\mu^*\mu\hat{\sigma}_+\,,
\end{aligned}
\end{equation}
and
\begin{equation}
\begin{aligned}
e^{i\hat{R}}
|e\rangle\langle e|e^{-i\hat{R}}
&=|e\rangle\langle e|
+i[\hat{R},|e\rangle\langle e|]
+\frac{i^2}{2!}
[\hat{R},
[\hat{R},|e\rangle\langle e|]]+\mathcal{O}
(|\mu|^3)\\
&\approx|e\rangle\langle e|
+i(\mu^*\hat{\sigma}_--\mu\hat{\sigma}_+)
-|\mu|^2\hat{\sigma}_z\\
&=
|e\rangle\langle e|
-\frac{\eta^*\Omega_a}{\delta_{al}}\hat{\sigma}_-
-\frac{\eta\Omega_a}{\delta_{al}}\hat{\sigma}_+
-|\frac{\eta\Omega_a}{\delta_{al}}|^2\hat{\sigma}_z\,.
\end{aligned}
\end{equation}
Note that, $\exp(i\hat{R})\hat{\sigma}_-\exp(-i\hat{R})
=[\exp(i\hat{R})\hat{\sigma}_+\exp(-i\hat{R})]^\dagger$, and the following commutators are used,
\begin{equation}
\begin{aligned}
&[\hat{R},\hat{\sigma}_z]=2(\mu^*\hat{\sigma}_--\mu\hat{\sigma}_+)\,,
\,\,\,\,\,\,\,\,
[\hat{R},|e\rangle\langle e|]=\mu^*\hat{\sigma}_--\mu\hat{\sigma}_+\,,\\
&[\hat{R},\hat{\sigma}_+]=-\mu^*\hat{\sigma}_z\,,\,\,\,\,\,\,\,\,
[\hat{R},\hat{\sigma}_-]=\mu\hat{\sigma}_z\,,\,\,\,\,\,\,\,\,
\hat{\sigma}_z=|e\rangle\langle e|-|g\rangle\langle g|\,.
\end{aligned}
\end{equation}

Following the above BCH formulas, Hamiltonian (16) is transformed to be
\begin{equation}
\begin{aligned}
\hat{H}_4
=&\hat{H}''_{\rm NCF}+\frac{\textbf{p}_a^2}{2m_a}+\frac{\textbf{p}'^2}{2m}
+\widetilde{H}(|\eta\Omega_a\mu|^2\hat{\sigma}_{\pm})+\hbar\delta_{al}|e\rangle\langle e|
+\hbar\frac{|\eta|^2\Omega_a^2}{\delta_{al}}\hat{\sigma}_z\,\,
\end{aligned}
\end{equation}
with
\begin{equation}
\begin{aligned}
e^{i\hat{R}}(\hat{H}''_{\rm NCF}+\frac{\textbf{p}_a^2}{2m_a}+\frac{\textbf{p}'^2}{2m})
e^{-i\hat{R}}\approx\hat{H}''_{\rm NCF}+\frac{\textbf{p}_a^2}{2m_a}+\frac{\textbf{p}'^2}{2m}\,.
\end{aligned}
\end{equation}
The last term in (20) is the standard Stark effect of atom. The term $\widetilde{H}(|\eta\Omega_a\mu|^2\hat{\sigma}_{\pm})$ refers to the high order small quantity $|\mu|^2$, and is negligible.
The unitary transformation $\exp(-i\hat{R})$, with $\hat{R}=\mu\hat{\sigma}_++\mu^*\hat{\sigma}_-$, made for producing Hamiltonian (20) would excite atom, but the probability is small (with  $|\mu|\rightarrow0$), so we list the approximate Eq.~(21).  Considering atom retaining on its internal ground state, then Hamiltonian (20) reduce to
\begin{equation}
\begin{aligned}
\hat{H}_4
=&\hat{H}''_{\rm NCF}+\frac{\textbf{p}_a^2}{2m_a}+\frac{\textbf{p}'^2}{2m}
-\hbar\Omega_{\rm eff}(\hat{x},\hat{x}_a)\,,
\end{aligned}
\end{equation}
with the effective Rabi frequency $\Omega_{\rm eff}(\hat{x},\hat{x}_a)=|\eta|^2\Omega_a^2/\delta_{al}$, which couples the external motions of nanopartcile and atom, because $\eta$ and $\Omega_{a}$ are associated with nanoparticle's position $x$ and atomic position $x_a$, repectively.

\subsection{The short time approximation}
We now take the last unitary transformation
\begin{equation}
\begin{aligned}
\hat{U}_{\rm {NA}}(t)=\exp\left[\frac{-it}{\hbar}\left(\frac{\textbf{p}_a^2}{2m_a}
+\frac{\textbf{p}^2}{2m}-\hbar\Omega_{\rm eff}(\hat{x},\hat{x}_a)\right)\right]
\end{aligned}
\end{equation}
to the total system.
The Hamiltonian (22) becomes
\begin{equation}
\begin{aligned}
\hat{H}_5(t)
=&\hat{H}'''_{\rm NCF}+\frac{\textbf{p}''^2}{2m}-\frac{\textbf{p}^2}{2m}\,,
\end{aligned}
\end{equation}
with $\hat{H}'''_{\rm NCF}=\hat{U}_{\rm {NA}}^\dagger(t)\hat{D}^\dagger(t)\exp(-i\delta_{c l}t\hat{a}^\dagger\hat{a})\hat{H}_{\rm NCF}\exp(i\delta_{c l}t\hat{a}^\dagger\hat{a})\hat{D}(t)\hat{U}_{\rm {NA}}(t)$ and $\hat{\textbf{p}}''=\hat{U}_{\rm {NA}}^\dagger(t)\hat{D}^\dagger(t)\hat{\textbf{p}}\hat{D}(t)\hat{U}_{\rm {NA}}$.
Recalling $\hat{H}_{\rm NCF}=\widetilde{H}_{\rm NF}+\hat{H}_{\rm CF}$, where $\widetilde{H}_{\rm NF}$ is the Hamiltonian of nanoparticle interacting with some other cavity modes and the free fields, and $\hat{H}_{\rm CF}$ the Hamiltonian of cavity modes interacting with the free fields. Therefore, the Hamiltonian $\hat{H}_5$ arise decoherence to the motional particles, because the operator $\hat{D}(t)$ of driving cavity is associated with nanoparticle's position and the bosonic operators of photon. The $\hat{U}_{\rm {NA}}(t)$ depends not only on nanoparticle's position but also on atomic position. Nevertheless, it is reasonable to assume $\hat{H}_5\rightarrow0$ within the weak driving optical system. This can be explained by the case of single mode cavity, i.e., $\hat{D}(t)\rightarrow1$ and $\hat{U}_{\rm {NA}}(t)\rightarrow\hat{U}_{0}(t)$ with $t\Omega_l\rightarrow0$. The $\hat{U}_0(t)=\exp(-it\hat{H}_K/\hbar)$ is the free evolution operator of the motional atom and nanoparticle when the input laser is absent, and $\hat{H}_K=\textbf{p}_a^2/(2m_a)+\textbf{p}/(2m)$ the kinetic energy of two particles.

According to the several unitary evolutions made above, the time evolution operator of the laser-manipulated cavity system can be written in a general form of
\begin{equation}
\begin{aligned}
\hat{U}_{L}(t)
&=\hat{D}(t)\hat{U}_{\rm {NA}}(t)\hat{U}_{5}(t)\,.
\end{aligned}
\end{equation}
The $\hat{U}_{\rm NA}(t)$ is the leading term about atom, and coupling with nanoparticle, so we use the subscript NA to mark it.
The $\hat{U}_{5}(t)$ is the time-evolution operator of Hamiltonian $\hat{H}_5(t)$.
Considering the laser pulse is sufficiently short, i.e., $t=\tau\rightarrow0$, then the evolution operator (25) can be approximately written as
\begin{equation}
\begin{aligned}
\hat{U}_{L}(\tau)
&=\hat{D}(\tau)\hat{U}_{\rm NA}(\tau)\hat{U}_{5}(\tau)\\
&\approx\hat{D}(\tau)\hat{U}_{5}(\tau)\hat{U}_{\rm NA}(\tau)\\
&\approx\hat{D}(\tau)\hat{U}_{5}(\tau)e^{\frac{-i\tau}{\hbar}\hat{H}_K}
e^{i\tau\Omega_{\rm eff}(\hat{x},\hat{x}_a)}\\
&\approx\hat{U}_0(\tau)\widetilde{U}(\tau)
e^{i\tau\Omega_{\rm eff}(\hat{x},\hat{x}_a)}\,,
\end{aligned}
\end{equation}
with a perturbation operator $\widetilde{U}(\tau)
=\hat{U}_0^\dagger(\tau)\hat{D}(\tau)\hat{U}_{5}(\tau)\hat{U}_0(\tau)$ of the cavity system.
In the second line of Eq.~(26), we have used the approximation $\hat{U}^\dagger_{5}(\tau)\hat{U}_{\rm NA}(\tau)\hat{U}_{5}(\tau)\approx\hat{U}_{\rm NA}(\tau)$. In the third line of Eq.~(26), we have used the approximation $\hat{U}_{\rm NA}(\tau)\approx\exp(-i\tau\hat{H}_K/\hbar)\exp(i\tau\Omega_{\rm eff})$, by neglecting the terms of $\tau^2[\Omega_{\rm eff},\hat{H}_K]/(2\hbar)$.
The present two approximations are both limited within the short time interaction, similar to that of the so-called Raman-Nath in atom optics~\cite{ZHU-ATOM}. Certainly, the short time interaction decreases also the desired signal of $\tau\Omega_{\rm eff}$, so we consider using atom interferometry to measure it. Once using atom as the detector, we can simply regard $\widetilde{U}(\tau)=1$, because $\tau\Omega_{\rm eff}$ is already a small quantity. As a consequence, we can directly use the leading term
\begin{equation}
\begin{aligned}
\hat{U}_{L}(\tau)
\approx\hat{U}_0(\tau)
e^{i\tau\Omega_{\rm eff}(\hat{x},\hat{x}_a)}
\end{aligned}
\end{equation}
to compute the Ramsey-Bord\'{e} atoms interferometer in Sec. III.

\subsection{The data estimations for the effective Rabi frequency}
According to Eq.~(14),  we rewrite the effective Rabi frequency as
\begin{equation}
\begin{aligned}
\Omega_{\rm eff}(\hat{x},\hat{x}_a)=\frac{|\eta|^2\Omega_a^2}{\delta_{al}}
=\Omega_{\rm effm}\cos^2(kx)\cos^2(kx_a)\,,
\end{aligned}
\end{equation}
with the nanoparticle's position $x$ and atomic position $x_a$. Here, $\Omega_{\rm effm}=\eta_{0}^2\Omega_{a0}^2/\delta_{al}$ is maximum coupling frequency, with
\begin{equation}
\begin{aligned}
\eta_{0}=
\frac{\epsilon_0\epsilon_c V_N\textbf{E}_c\cdot\textbf{E}_l}
{2\hbar(\omega_c-\omega_l-\Omega_c)}\approx\frac{\epsilon_0\epsilon_c V_N\textbf{E}_c\cdot\textbf{E}_l}
{2\hbar(\omega_c-\omega_l)}\,.
\end{aligned}
\end{equation}
The weak coupling $\Omega_c\ll\omega_c-\omega_l$ is considered, and where $\Omega_c=\Omega_{c0}\cos^2(kx)$ is given by Eq.~(6).
We use the two-photon Rabi frequency $\Omega_{ c0}=\epsilon_0\epsilon_cV_{N}\textbf{E}_c\cdot\textbf{E}_c/(2\hbar)
=\epsilon_cV_{N}\omega_c/(4V_c)$ of nanoparticle-cavity interaction as the criterion to take the numerical estimations, and where $\textbf{E}_c=\textbf{e}_c\sqrt{\hbar\omega_c/(2\epsilon_0V_c)}$ is the vacuum fluctuation of electric field strength of cavity mode $\omega_c$. In terms of $\Omega_{ c0}$, Eq.~(29) reads
\begin{equation}
\begin{aligned}
\eta_{0}=\frac{\Omega_{c0}}{\omega_c-\omega_l}
\frac{(\textbf{e}_c\cdot\textbf{e}_l)E_l}{E_{c}}\,,
\end{aligned}
\end{equation}
and where $\textbf{e}_c$ and $\textbf{e}_l$ are the unit vectors of cavity field and the laser field, respectively.

Considering the wavelength of input laser is approximate $\lambda=780$ nm, and the effective volume of cavity is $V_c\approx\pi(w/2)^2L$, with the waist $w=40\,\,\mu{\rm m}$, and the length $L=1$~cm.
Here the $\lambda=780$ nm is different from 1064 nm in the usual nanoparticle experiment~\cite{EXP-Science}, because we need a light which is not so large detuning with the ${\rm D}_2$ transition of ${\rm Rb}$ atom~\cite{D2,102,g12} for generating the considerable effects. Furthermore, considering a volume $V_N=4\pi R^3/3$ of nanoparticle with radius $R=150$~nm and the relative dielectric constant $\epsilon_r=2.1$ ~\cite{EXP-NJP2013}, we have the  two-photon Rabi frequency of nanoparticle-cavity interaction, $\Omega_{ c0}\approx1.4$~MHz with  $\epsilon_c=3(\epsilon_r-1)/(\epsilon_r+2)$.
The single photon Rabi frequency of atom-cavity interaction is computed by $\Omega_{a0}=\textbf{d}_a\cdot\textbf{E}_c/\hbar
=(\textbf{d}_a\cdot\textbf{e}_c)
\sqrt{\omega_c/(2\epsilon_0V_c\hbar)}\approx12$ MHz, with the atomic transition dipole moment $(\textbf{d}_a\cdot\textbf{e}_c)\approx 3.6\times10^{-29}\,\,{\rm C}\cdot{\rm m}$~\cite{D2}.

The above $\Omega_{\rm c0}\approx1.4$~MHz and $\Omega_{a0}\approx12$ MHz are too large because the waist $w=40\,\,\mu{\rm m}$ of cavity is small (used originally in the nanoparticle cooling experiments)~\cite{EXP-NJP2013}. This would be not practical for generating the coupling between the freely flying atom and nanoparticle (as they should be inside the cavity at the same time). A more larger cavity is needed, for example, $w=1\,\,{\rm mm}$ and $L=2$~cm. As a result, we have the two-photon Rabi frequency $\Omega_{c0}\approx1.1$~kHz of nanoparticle and the single-photon Rabi frequency $\Omega_{a0}\approx0.33$ MHz of atom. We further consider the detuning $\delta_{al}\approx10\times\Omega_{a0}\approx3.3$ MHz between atom and the external laser, and then the two-photon Rabi frequency of atom is $\Omega_{a0}^2/\delta_{al}\approx33$ KHz. Moreover, considering
$\eta_0=5$ with $\omega_c-\omega_l\approx(\textbf{e}_c\cdot\textbf{e}_lE_l/E_c)\times0.22$~KHz and $(\textbf{e}_c\cdot\textbf{e}_lE_l/E_c)>1.5\times10^3$, we have the value of the effective Rabi frequency $\Omega_{\rm effm}=\eta_{0}^2\Omega_{a0}^2/\delta_{al}\approx0.83$ MHz.
The Raman-Nath approximation works within the regime $kv\tau\ll1$ with $v$ being the characteristic velocity of particle. The velocities of cold nanoparticle and cold atom are on the orders of $v=13\,\mu{\rm m}/{\rm s}$~\cite{EXP-Science} and $v_a=2\,{\rm mm}/{\rm s}$~\cite{KD-E}, respectively. As a consequence, $kv\tau=2\pi v\tau/\lambda\approx1.05\times10^{-5}\lll1$ and $kv_a\tau=2\pi v_a\tau/\lambda\approx1.6\times10^{-3}\ll1$ with $\lambda=780$~nm and a short time $\tau=0.1\,\,\mu{\rm s}$. Using such an interaction time, we estimate the coupling strength $\tau\Omega_{\rm effm}=0.083$.

\section{The Ramsey-Bord\'{e} atom interferometer}
Following Eq.~(27), we expand the atomic term as
\begin{equation}
\begin{aligned}
e^{i\tau\Omega_{\rm eff}(\hat{x},\hat{x}_a)}&\approx1+i\xi\cos^2(k\hat{x})
\left(2+e^{i2k\hat{x}_a}+e^{-i2k\hat{x}_a}\right)+\mathcal{O}(\xi^2)
\,,
\end{aligned}
\end{equation}
by neglecting the probability-amplitude of nonlinear $\xi^2=(\Omega_{\rm effm}\tau/4)^2$ throughout the paper. This operator can be regarded as a coupled KD scattering of atom and nanoparticle, and where $\exp(\pm 2k\hat{x}_a)$ changes $x$ directional momentum of atom in terms of
$\pm 2\hbar k$, resulting in three distinguishable paths, because $\hbar k\approx m_a\times 5.8\,\,{\rm mm}\cdot{\rm s}^{-1}$ is larger than the initial momentum uncertainty of cold atom~\cite{KD-E}. This is why atoms can be used to make a matter wave interferometer, but nanoparticle is difficult.

\subsection{The freely expanding Gaussian wave packet of nanoparticle}
According to Eq.~(31), perhaps the most direct way of nanoparticle detections is to observe how many atoms are scattered.
The probability of finding the scattered atoms reads
\begin{equation}
\begin{aligned}
\xi^2\langle N|\cos^4(kx)|N\rangle\approx\frac{3\xi^2}{8}+\frac{\xi^2}{4}\langle N|e^{i2kx}|N\rangle+\frac{\xi^2}{16}\langle N|e^{i4kx}|N\rangle+{\rm c.c.}\,,
\end{aligned}
\end{equation}
where the ${\rm c.c.}$ is the conjugate complex number of all the non-real numbers in the right hand of the equation.
For the state of nanoparticle, we consider the simplest case, that the freely expanding Gaussian wave packet along horizontal direction,
\begin{equation}
\begin{aligned}
|N\rangle=e^{\frac{-it}{\hbar}\frac{\hat{p}^2}{2m}}|G\rangle,
\end{aligned}
\end{equation}
with
\begin{equation}
\begin{aligned}
|G\rangle=\int_{-\infty}^{+\infty}\phi(p)|p\rangle {\rm d}p
=(2\pi\Delta_p^2)^{-1/4}\int_{-\infty}^{+\infty}
e^{\frac{-p^2}{4\Delta_p^2}}|p\rangle {\rm d}p\,,
\end{aligned}
\end{equation}
and where $\Delta_p$ is the momentum uncertainty of initial nanoparticles. This state can be generated by a released nanoparticle which was initially confined as a harmonic oscillator in the optical trap and cooled at the motional quantum ground state~\cite{EXP-Science}.
Using the Heisenberg operator of uniform rectilinear motion,
\begin{equation}
\begin{aligned}
e^{\frac{it}{\hbar}\frac{\hat{p}^2}{2m}}\hat{x}
e^{\frac{-it}{\hbar}\frac{\hat{p}^2}{2m}}
=\hat{x}+\frac{\hat{p}}{m}t=\hat{x}+\hat{v}t\,,
\end{aligned}
\end{equation}
we write one of the terms in Eq.~(32) as
\begin{equation}
\begin{aligned}
\langle N|e^{ink\hat{x}}|N\rangle
=\langle G|e^{ink(\hat{x}+\hat{v}t)}|G\rangle=e^{\frac{i\hbar n^2 k^2 t}{2m}}\langle G|e^{ink\hat{x}}e^{ink\hat{v}t}|G\rangle\,,
\end{aligned}
\end{equation}
with $n=2,4$. The phase $\hbar n^2 k^2 t/(2m)=n^2kv_kt/2$, with the single photon recoil velocity $v_k=\hbar k/m$ of nanoparticle, should be the nonclassical one, because it came from the canonical quantization $[\hat{x},\hat{v}]=i\hbar/m$ of nanoparticle.

Unfortunately, the phase $\hbar n^2 k^2 t/(2m)$ is exactly eliminated by the following Gaussian integral
\begin{equation}
\begin{aligned}
\langle G|e^{ink\hat{x}}e^{ink\hat{v}t}|G\rangle
&=\int_{-\infty}^{+\infty} e^{\frac{inkpt}{m}}\phi^*(p+n\hbar k)\phi(p){\rm d} p\\
&=(2\pi\Delta_p^2)^{-1/2}
\int_{-\infty}^{+\infty}e^{\frac{inkpt}{m}}
e^{\frac{-p^2}{4\Delta_p^2}}
e^{\frac{-(p+n\hbar k)^2}{4\Delta_p^2}}{\rm d} p\\
&=\exp\left(-\frac{n^2\hbar^2k^2}{8\Delta_p^2}\right)
\exp\left[-\frac{1}{2}\left(\frac{nk\Delta_pt}{m}\right)^2\right]
\exp\left(-i\frac{\hbar n^2k^2t}{2m}\right)\,.
\end{aligned}
\end{equation}
Thus, the $\cos^n(k\hat{x})$ applied to measure the freely expanding Gaussian wave packet does not produce nonclassical phenomena. Nevertheless, the $\cos^n(k\hat{x})$ may be useful for measuring the momentum uncertainty of nanoparticle.
Numerically, considering the momentum uncertainty is that of ground state harmonic oscillator~\cite{EXP-Science}, i.e., $\Delta_p=m\times13\,\,\mu{\rm m}\cdot{\rm s}^{-1}$ introduced in Sec. I., we find
$\exp\left[-\hbar^2n^2k^2/(8\Delta_p^2)\right]\approx1$, because the single photon recoiled momentum $\hbar k\approx m\times5\,\,{\rm nm}\cdot{\rm s}^{-1}$ is far smaller than
the initial momentum uncertainty $\Delta_p$ of nanoparticle.  The value of $\exp\left[-(nk\Delta_pt/m)^2/2\right]$ is considerable, if $\Delta_pt/m<1/(2k)=\lambda/(4\pi)$, i.e., $t<5$~ms.

One may suppose a $x$ directional `` gravity'' acceleration $g_x$, then the familiar phase $nkg_xt^2/2$ in Mach-Zehnder atom interferometer~\cite{ChuG} is obtained by replacing Heisenberg's operator (35) with $\hat{x}+\hat{v}t+g_xt^2/2$. However, this phase should be not regarded as the nonclassical effect of nanoparticle.
The Eq.~(37) indicates that a long time evolution of Gaussian wave packet results in a small signal due to the spatial symmetry of function $\cos^n(kx)$. This means that a more smaller momentum uncertainty $\Delta_p$ of the initial nanoparticle is needed. One may use the locality of light spot to measure nanoparticle. However, the waist of laser is much larger than its wave length, so the resolution ratio of particle's position would be not significantly high by using this approach.
In this paper, we focus on the effects from the short wave length $\lambda=2\pi/k$ of light. Although the above measurement does not generate the signal about nanoparticle's motions (after a long time free evolution of Gaussian wave packet), the first term $3\xi^2/8$ in Eq.~(32) is still useful for extracting the nonclassical signals of nanoparticle in the following Ramsey-Bord\'{e} atom interferometer.

\begin{figure}[tbp]
\includegraphics[width=10cm]{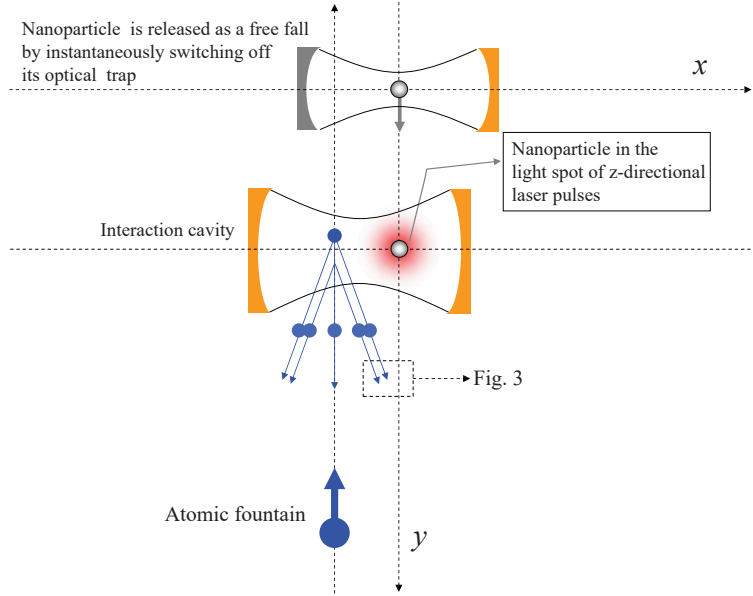}
\caption{Sketch for the conceptual experiment. Inside the interaction cavity, nanoparticle is illuminated by two $z$-directional laser pulses at $t_1$ and $t_2$, generates twice atomic KD scattering (31) to realize the $+x$ directional Ramsey-Bord\'{e} atom interferometer showed in Fig.~3. Inside the interaction cavity, the nanoparticle's velocity $v_{y}$ and atomic velocity $v_{ay}$ are not so large that the particles will pass through the cavity before the two laser pulses are completed. For example, $v_{ay}<v_{y}<w/\delta t\approx1\,\,{\rm m\cdot s}^{-1}$, with $w\approx1\,\,{\rm mm}$ being the waists of cavity and the laser beam, and $\delta t=t_2-t_1\approx1$~ms the time interval between two laser pulses. The $-x$ direction scattered atoms are not used to make the Ramsey-Bord\'{e} atom interferometer, but used to measure the values of $\xi_j$ based on Eq.~(61), allowing us to extract the desired nonclassical signal in Ramsey-Bord\'{e} atom interferometer.}
\end{figure}
\begin{figure}[tbp]
\includegraphics[width=12cm]{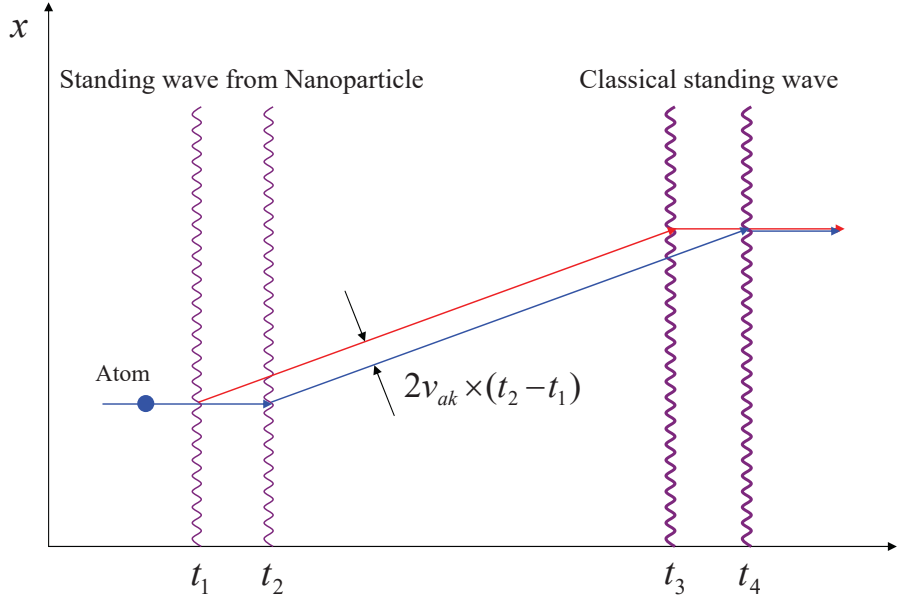}
\caption{Sketch for the $+x$ directional Ramsey-Bord\'{e} atom interferometer, with $t_2-t_1=t_4-t_3$. Two weak standing waves at $t_1$ and $t_2$ are due to the laser-illuminated nanoparticle, which split the $x$ directional paths of atom. Two classical laser pulses at $t_3$ and $t_4$ are used to recombine atomic paths for completing the matter wave interferometer. The large distance $2v_{ak}\times(t_2-t_1)$ between the red and blue lines is the key to generate the visible atom interference, for example, $2v_{ak}\times(t_2-t_1)\approx12\,\,\mu{\rm m}$ may be large enough, with $t_2-t_1\approx1$ ms and the single photon recoil velocity $v_{ak}\approx5.8\,\,{\rm mm}\cdot{\rm s}^{-1}$ of atom.}
\end{figure}

\subsection{The time-evolution operator approach for Ramsey-Bord\'{e} atom interferometer}
We design a symmetric Ramsey-Bord\'{e} atom interferometer~\cite{SRB} to measure the small nonclassicality of nanoparticle. This atom interferometer is made by using twice interaction (31) at times $t_1$ and $t_2$ to split atomic path along horizontal $x$ direction, and consequently, using two classical laser pulses at times $t_3$ and $t_4$ to recombine atomic paths, as shown in Fig.~2 and Fig.~3.

The KD scatterings for making atom interferometers will generate many distinguishable paths of freely falling atom~\cite{102,g12}. Nevertheless, we can consider just only two selected paths of atoms, because the atoms moving along the other paths do not arrive at the small detection zone of atoms.
We use $\alpha$ and $\beta$ to denote the probability-amplitudes of atoms which came from the two selected paths, i.e., the red line and blue line in Fig.~3, and write the final state as
\begin{equation}
\begin{aligned}
|\psi\rangle
&=\alpha |{\rm Red}\rangle+\beta|{\rm Blue}\rangle+\cdots\\
&=\hat{R}|A(t_1)\rangle|N(t_1)\rangle
+\hat{B}|A(t_1)\rangle|N(t_1)\rangle+\cdots\,.
\end{aligned}
\end{equation}
The $|{\rm Red}\rangle$ and $|{\rm Blue}\rangle$ are the states of the total system, with atom moving along the red and the blue lines, respectively. The two states are further denoted by the path-operators $\hat{R}$ and $\hat{B}$, with the initial state $|A(t_1)\rangle|N(t_1)\rangle$ at $t_1$, i.e., the moment of the first laser pulse of Ramsey-Bord\'{e} atom interferometer.

According to Fig.~3, we can write the path-operators as
\begin{equation}
\begin{aligned}
\hat{R}=&i\xi_1\alpha_{l}
\hat{U}_{0}(t_4-t_3)
e^{-i2k\hat{x}_a}
\hat{U}_{0}(t_3-t_1)
\cos^2(k\hat{x})e^{i2k\hat{x}_a}\,,
\end{aligned}
\end{equation}
and
\begin{equation}
\begin{aligned}
\hat{B}=i\xi_2\beta_{l}
e^{-i2k\hat{x}_a}
\hat{U}_{0}(t_4-t_2)
\cos^2(k\hat{x})e^{i2k\hat{x}_a}\hat{U}_{0}(t_2-t_1)\,.
\end{aligned}
\end{equation}
Here, the probability-amplitudes $\alpha_{l}$ and $\beta_l$ are due to the classical laser pulses used for recombining atomic paths. The $\hat{U}_0(t)=\hat{U}_{\rm N0}(t)\hat{U}_{\rm A0}(t)$ is the free evolution operator of particles (with negligible interaction energy between them), and where $\hat{U}_{\rm N0}(t)$ is the free evolution operator of nanoparticle. The scattering operator $\exp(\pm i2k\hat{x}_a)$ changes atomic momentum instantaneously, resulting in two distinguishable paths due to the free evolution operator $\hat{U}_{\rm A0}(t)$ of atom.
The $i\xi_j
\cos^2(k\hat{x})$, with $j=1,2$, is due to the Eq.~(31), and can be also regarded as the probability-amplitudes of atomic scattering, if nanoparticle is the classical one.

Using the property of unitary operator $\hat{U}_{0}^\dagger(t)\hat{U}_{0}(t)=\hat{U}_{0}(t)\hat{U}_{0}^\dagger(t)=1$ and the commutator $[\hat{U}_{\rm A0}(t),\hat{U}_{\rm N0}(t)]=0$ of atom and nanoparticle, we further write the states (39) and (40) as
\begin{equation}
\begin{aligned}
\hat{R}=&i\xi_1\alpha_{l}
\hat{U}_{0}(t_4-t_1)
e^{-i2k\hat{x}_a(t_3-t_1)}e^{i2k\hat{x}_a}
\cos^2(k\hat{x})\,,
\end{aligned}
\end{equation}
and
\begin{equation}
\begin{aligned}
\hat{B}=&i\xi_2\beta_{l}\hat{U}_{0}(t_4-t_1)
e^{-i2k\hat{x}_a(t_4-t_1)}e^{i2k\hat{x}_a(t_2-t_1)}
\cos^2[k\hat{x}(t_2-t_1)]\,.
\end{aligned}
\end{equation}
Here, $\hat{x}_a(t)
=\hat{x}_a+\hat{v}_{a}t$ and $\hat{x}(t)
=\hat{x}+\hat{v}t$ are the Heisenberg's operators of atom and nanoparticle moving along the horizontal $x$ directions, see also Eq.~(35).

Eq.~(41) and (42) can be written in more clear, i.e., $\hat{R}=\hat{R}_{\rm A}\hat{R}_{\rm N}$ and $\hat{B}=\hat{B}_{\rm A}\hat{B}_{\rm N}$, with atomic parts $\hat{R}_{\rm A}$, $\hat{B}_{\rm A}$, and nanoparticle's parts $\hat{R}_{\rm N}$ and $\hat{B}_{\rm N}$. The four operators are given by follows,
\begin{equation}
\begin{aligned}
\hat{R}_{\rm A}=&
\hat{U}_{A0}(t_4-t_1)
e^{-i2k\hat{x}_a(t_3-t_1)}e^{i2k\hat{x}_a}\,,
\end{aligned}
\end{equation}
\begin{equation}
\begin{aligned}
\hat{B}_{\rm A}=&\hat{U}_{\rm A0}(t_4-t_1)
e^{-i2k\hat{x}_a(t_4-t_1)}e^{i2k\hat{x}_a(t_2-t_1)}\,,
\end{aligned}
\end{equation}
\begin{equation}
\begin{aligned}
\hat{R}_{\rm N}=&i\alpha_{l}\xi_1
\hat{U}_{\rm N0}(t_4-t_1)
\cos^2(k\hat{x})\,,
\end{aligned}
\end{equation}
\begin{equation}
\begin{aligned}
\hat{B}_{\rm N}=&i\beta_{l}\xi_2\hat{U}_{\rm N0}(t_4-t_1)
\cos^2[k\hat{x}(t_2-t_1)]\,.
\end{aligned}
\end{equation}
Considering $t_3-t_1=t_4-t_2=T$, i.e., $t_4-t_3=t_2-t_1$, the atomic operators reduce to
\begin{equation}
\begin{aligned}
\hat{R}_{\rm A}=\hat{B}_{\rm A}=
\hat{U}_{A0}(t_4-t_1)e^{-i2\hbar k^2T/m_a}
e^{-i2k\hat{v}_aT}\,,
\end{aligned}
\end{equation}
and here the so-called Zassenhaus formula~\cite{Zassenhaus-2012,Zassenhaus-1967} is used,
\begin{equation}
\begin{aligned}
e^{i2k[\hat{x}_a(t_i)-\hat{x}_a(t_j)]}
=e^{-i2k\hat{x}_a(t_j)}
e^{i2k\hat{x}_a(t_i)}e^{-2k^2[\hat{x}_a(t_j),\hat{x}_a(t_i)]}\,,
\end{aligned}
\end{equation}
with  $[\hat{x}_a(t_j),\hat{x}_a(t_i)]=i\hbar (t_i-t_j)/m_a$.

According to Eq.~(47), the superposition state (38) reads
\begin{equation}
\begin{aligned}
|\psi\rangle
&=\left[\hat{R}_{\rm N}|N(t_1)\rangle
+\hat{B}_{\rm N}|N(t_1)\rangle\right]|A(t_4)\rangle+\cdots\,,
\end{aligned}
\end{equation}
with atomic state $|A(t_4)\rangle
=\hat{R}_{\rm A}|A(t_1)\rangle=\hat{B}_{\rm A}|A(t_1)\rangle$ at $t_4$.  As a result, the probability density of atom
reads $P_a(x_a)=
|\langle x_a|\psi\rangle|^2
=|\langle x_a|A(t_4)\rangle|^2\times P$, with
\begin{equation}
\begin{aligned}
P&=\langle N(t_1)|(\hat{R}_{\rm N}+\hat{B}_{\rm N})^\dagger(\hat{R}_{\rm N}+\hat{B}_{\rm N})|N(t_1)\rangle\,.
\end{aligned}
\end{equation}
The $P$ is the desired signal.
According to Eqs.~(45) and (46), we have
\begin{equation}
\begin{aligned}
\hat{R}^\dagger_{\rm N}\hat{R}_{\rm N}\approx|\alpha_{l}|^2\xi_1^2\cos^4(k\hat{x})+\mathcal{O}(\xi^3)\,,
\end{aligned}
\end{equation}
\begin{equation}
\begin{aligned}
\hat{B}^\dagger_{\rm N}\hat{B}_{\rm N}&\approx|\beta_{l}|^2\xi_2^2\cos^4[k\hat{x}(t_2-t_1)]+\mathcal{O}(\xi^3)\,,
\end{aligned}
\end{equation}
\begin{equation}
\begin{aligned}
\hat{R}^\dagger_{\rm N}\hat{B}_{\rm N}&\approx\alpha_l^*\beta_l\xi_1\xi_2
\cos^2(k\hat{x})\cos^2[k\hat{x}(t_2-t_1)]+\mathcal{O}(\xi^3)\,.
\end{aligned}
\end{equation}
The higher orders than $\xi^2$ are neglected.

\subsection{The long time expanding Gaussian wave packet of nanoparticle}
In order to compute Eq.~(50), we should know the `` initial'' state of nanoparticle at $t_1$.
The nanoparticle's state at $t_0$ is that at the moment of nanoparticle releasing from the optical harmonic oscillator potential, which can be regarded as a Gaussian wave packet $|N(t_0)\rangle=|G\rangle$. Thus, $|N(t_1)\rangle=\exp[-i(t_1-t_0)p^2/(2m\hbar)]|G\rangle$, if nanoparticle in stage $t_1-t_0$ is really a free fall, and therefore Eq.~(50) becomes
\begin{equation}
\begin{aligned}
P=&|\alpha_{l}|^2\xi_1^2\langle G|\cos^4[k\hat{x}(\Delta t_1)]|G\rangle+|\beta_{l}|^2\xi_2^2\langle G|\cos^4[k\hat{x}(\Delta t_2)]|G\rangle\\
&+\alpha_l^*\beta_l\xi_1\xi_2
\langle G|\cos^2[k\hat{x}(\Delta t_1)]\cos^2[k\hat{x}(\Delta t_2)]|G\rangle
+{\rm c.c.}\,,
\end{aligned}
\end{equation}
with $\Delta t_j=t_j-t_0$ and $j=1,2$.

In order to compute Eq.~(54), we introduce the following Gaussian integral,
\begin{equation}
\begin{aligned}
\langle G|e^{in_1k\hat{x}}e^{in_2k\hat{v}\Delta t_j}|G\rangle
=&\exp\left(-\frac{\hbar^2n_1^2k^2}{8\Delta_p^2}\right)
\exp\left[-\frac{1}{2}\left(\frac{n_2k\Delta_p\Delta t_j}{m}\right)^2\right]\\
&\times
\exp\left(-i\frac{ n_1n_2kv_k\Delta t_j}{2}\right)\,,
\end{aligned}
\end{equation}
with $n_1$ and $n_2$ being arbitrary real numbers, and $v_k=\hbar k/m$ the single photon recoil velocity of nanoparticle (not atom). This integral is a generalization of Eq.~(37), and where $\Delta_p=m\times13\,\,\mu{\rm m}\cdot{\rm s}^{-1}$ is the momentum uncertainty of the initially cooled nanoparticle, and $k=2\pi/(780\,{\rm nm})$ the wave number of the cavity mode.

Considering a long time free fall before the Ramsey-Bord\'{e} atom interferometer, e.g., $t_1-t_0\approx0.1$ s, and a short time between the laser pulses of Ramsey-Bord\'{e} atom interferometer, e.g., $\delta t=t_2-t_1=1$~ms, then $n_2\Delta_p\Delta t_j/m\gg1/(2k)$ with $n_2=2,4$, and the
Gaussian integral (55) is approximately equal to zero, see the discussion below Eq.~(37).
This greatly reduces the results of Eq.~(54), as follows,
\begin{equation}
\begin{aligned}
\langle G|\cos^4[k\hat{x}(\Delta t_1)]|G\rangle
\approx\langle G|\cos^4[k\hat{x}(\Delta t_2)]|G\rangle\approx\frac{3}{8}\,,
\end{aligned}
\end{equation}
and
\begin{equation}
\begin{aligned}
&\langle G|\cos^2[k\hat{x}(\Delta t_1)]\cos^2[k\hat{x}(\Delta t_2)]
|G\rangle\\
&\approx\frac{1}{4}+\frac{1}{4}\langle G|\cos[2k\hat{x}(\Delta t_1)]\cos[2k\hat{x}(\Delta t_2)]|G\rangle\\
&\approx
\frac{1}{4}+\frac{1}{16}\langle G|
\left[e^{i2k\hat{x}(\Delta t_1)}e^{-i2k\hat{x}(\Delta t_2)}
+e^{-2ik\hat{x}(\Delta t_1)}e^{i2k\hat{x}(\Delta t_2)}\right]|G\rangle\\
&=\frac{1}{4}+\frac{1}{16}e^{i\theta_q}
\langle G|\left[e^{-i2k\hat{v}(t_2-t_1)}
+e^{2ik\hat{v}(t_2-t_1)}\right]|G\rangle\\
&=\frac{1}{4}+\frac{1}{8}e^{i\theta_q}
G(\delta t)\,.
\end{aligned}
\end{equation}

In Eq.~(57), the Gaussian function
\begin{equation}
\begin{aligned}
G(\delta t)=\exp\left[-\frac{1}{2}
\left(\frac{2k\Delta_p\delta t}{m}\right)^2\right]\,,
\end{aligned}
\end{equation}
and the phase
\begin{equation}
\begin{aligned}
\theta_q=\frac{2\hbar k^2\delta t}{m}=2kv_k\delta t\,,
\end{aligned}
\end{equation}
are the signals of motional nanoparticle.
Specially, the phase $\theta_q$ is the quantum mechanical one
because it is generated by
the commutators
$[\hat{x}(\Delta t_1),\hat{x}(\Delta t_2)]
=i\hbar (t_2-t_1)/m$ of nanoparticle, and where $v_k=\hbar k/m$ is the single photon recoil velocity of nanoparticle.
According Eqs.~(56) and (57), the probability (54) is solved,
\begin{equation}
\begin{aligned}
P=\frac{3(\xi_1^2|\alpha_l|^2+\xi_2^2|\beta_l|^2)}{8}
+\frac{\xi_1\xi_2\alpha_l^*\beta_l}{4}\left[1
+\frac{1}{2}e^{i\theta_q}G(\delta t)\right]+{\rm c.c.}\,.
\end{aligned}
\end{equation}
Considering the time-interval $\delta t=t_2-t_1=1.2\,\,{\rm ms}$, the term $G(\delta t)\approx1$, and the phase $\theta_q=2kv_k\delta t
\approx10^{-4}$, with $v_k=\hbar k/m\approx5\,\,{\rm nm}\cdot{\rm s}^{-1}$, $k=2\pi/(780\,{\rm nm})$, and the mass $m=10^8$~amu of nanoparticle. The phase $\theta_q$ can be enlarged by increasing the time interval $\delta t=t_2-t_1$ between the two laser pulses. However, a considerable value of Gaussian function $G(\delta t)$ needs the small momentum uncertainty $\Delta p$ of initial nanoparticle, i.e., a more colder nanoparticle is much-needed. Using the data of experiments~\cite{EXP-Science}, the $\delta t$ is limited within $5$~ms, see the explanation below Eq.~(37).
In addition to the interest of $\theta_q$, the Eq.~(60) can be useful for testing the Gaussian function $G(\delta t)$, i.e., the initial momentum uncertainty of nanoparticle. Moreover, the general Eq.~(50) of the present Ramsey-Bord\'{e} atom interferometer could be also useful for measuring the non-Gaussian state at $t_1$, prepared by the strong KD scattering of nanoparticle in the classical laser pulse before $t_1$~\cite{NC2021,free-N-KD}.

\subsection{Discussion}
On the $\theta_q$ detection, the parameter $\xi_j$ has great uncertainty due to the imprecise positions of atom and nanoparticle relative to the waists of external laser and the cavity. Thus, the releasing times of particles and their initial velocities are needed to be exactly controlled. According to the symmetry of KD scattering, the $-x$ direction scattered atoms in Fig.~2 are also related to the parameter $\xi_j$. The probability of finding those atoms reads
\begin{equation}
\begin{aligned}
P_{\rm ref}&=\frac{3(\xi_1^2+\xi_2^2)}{8}\,,
\end{aligned}
\end{equation}
and can be measured before the $+x$ directional Ramsey-Bord\'{e} atom interferometer is completed. The probability (61) can be used as a reference to select the signal of $\theta_q$ in Eq.~(60).
Therefore, the $\alpha_l$ and $\beta_l$ become the main noises source due to the imperfect laser pulses at $t_3$ and $t_4$, as well as that in the usual atom interferometers.
In addition to the noises of laser manipulations, there exists environmental noises in the free fall stage of nanoparticle due to the collisions with residual gas particles and the blackbody radiation. This issue has been well studied by Romero-Isart et al.~\cite{Romero-Isart-collapse}. In principle, the environmental noises can be suppressed by using the ultrahigh vacuum chamber with the cryogenic temperatures~\cite{Romero-Isart-Science}. Launching satellite, for example the MAQRO project~\cite{MAQRO2012,MAQRO2015,NC2021}, should be the most effective way to realize the long time free fall and suppress the environmental noises.
Throughout the paper, we consider just only the single nanoparticle, so need the nanoparticle reloading. Similar to the magneto-optical trap in cold atoms system, the nanoparticle reloading could be realized by using the optical tweezer together with the gradient electric or magnetic fields~\cite{AN-Paul,CONTEMPORARY-PHYSICS,Romero-Isart-Science}.

Finally, we discuss the question of many atoms being simultaneously inside the cavity. According to the single atom scattering operator~(27), the operators of $n$ atoms scattered by two laser pulses can be written as $\hat{U}_{L1}=\Pi_{j=1}^n\exp(i\hat{\xi}_{j})$ and $\hat{U}_{L2}=\Pi_{j=1}^n\exp(i\hat{\xi}'_{j})$,
with $\hat{\xi}_{j}=4\xi_1\cos^2(k\hat{x})
\cos^2(k\hat{x}_{aj})$ and $\hat{\xi}'_{j}=4\xi_2\cos^2(k\hat{x})
\cos^2(k\hat{x}_{aj})$. The $\hat{x}_{aj}$ is the horizontal position of $j$-th atom. After the two scatterings (at $t_1$ and $t_2$), the state is prepared in
\begin{equation}
\begin{aligned}
|\psi(t_2)\rangle&=\hat{U}_{L2}
\hat{U}_0(t_2-t_1)\hat{U}_{L1}|\psi(t_1)\rangle=
\hat{U}_0(t_2-t_1)|\psi'(t_1)\rangle\,,
\end{aligned}
\end{equation}
with $\hat{U}_0(t)=\hat{U}_{N0}(t)\Pi_{j=1}^n\hat{U}_{j0}(t)$ being the free evolution operator of the nanoparticle and the many atoms. In this presentation, the `` initial'' state reads
\begin{equation}
\begin{aligned}
|\psi'(t_1)\rangle&=\hat{U}_{L2}(\delta t)\hat{U}_{L1}
|\psi(t_1)\rangle\\
&=[e^{i\hat{\xi}'_{1}(\delta t)}
e^{i\hat{\xi}'_{2}(\delta t)}
\cdots e^{i\hat{\xi}'_{n}(\delta t)}]
[e^{i\hat{\xi}_{1}}
e^{i\hat{\xi}_{2}}
\cdots e^{i\hat{\xi}_{n}}]|\psi(t_1)\rangle\\
&\approx[e^{i\hat{\xi}'_{1}(\delta t)}e^{i\hat{\xi}_{1}}]
[e^{i\hat{\xi}'_{2}(\delta t)}e^{i\hat{\xi}_{2}}]
\cdots [e^{i\hat{\xi}'_{n}(\delta t)}
e^{i\hat{\xi}_{n}}]|\psi(t_1)\rangle\,,
\end{aligned}
\end{equation}
with $\hat{\xi}'_j(\delta t)=\hat{U}_0^\dagger(\delta t)\hat{\xi}'_j\hat{U}_0(\delta t)$ and $\delta t=t_2-t_1$, and where
\begin{equation}
\begin{aligned}
|\psi(t_1)\rangle=
|A_1(t_1)\rangle|A_2(t_1)\rangle\cdots|A_n(t_1)\rangle|N(t_1)\rangle\,,
\end{aligned}
\end{equation}
is the state of $n+1$ particles at $t_1$, i.e., $n$ atoms and one nanoparticle.

In the last line of Eq.~(63), we have used the approximate Zassenhaus formula
\begin{equation}
\begin{aligned}
e^{i\hat{\xi}'_{i}(\delta t)}e^{i\hat{\xi}_{j}}
\approx e^{i\hat{c}_{ij}} e^{i\hat{\xi}_{j}}
e^{i\hat{\xi}'_{i}(\delta t)}\approx e^{i\hat{\xi}_{j}}
e^{i\hat{\xi}'_{i}(\delta t)}\,,
\end{aligned}
\end{equation}
and where
\begin{equation}
\begin{aligned}
\hat{c}_{ij}=i16\xi_1\xi_2
[\cos^2(k\hat{x}(\delta t)),\cos^2(k\hat{x})]\cos^2(k\hat{x}_{ai}(\delta t))\cos^2(k\hat{x}_{aj})
\end{aligned}
\end{equation}
is due to nanoparticle's commutator $[\cos^2(k\hat{x}(\delta t)),\cos^2(k\hat{x})]
=[\cos^2(k\hat{x}+k\hat{v}\delta t),\cos^2(k\hat{x})]\neq0$, which generates the atom-atom coupling. However, the probability-amplitudes are related to the quadratic small quantity $\xi^2$. In this paper, the probability-amplitudes are limited within the linear $\xi$, so we can neglect the terms of $\hat{c}_{ij}$ in the last line of Eq.~(63).

Following Eq.~(63), the prepared state by nanoparticle light sources can be written as
\begin{equation}
\begin{aligned}
|\psi'(t_1)=&|A'_1(t_1)\rangle|A'_2(t_1)
\rangle\cdots|A'_n(t_1)\rangle|N(t_1)\rangle\,,
\end{aligned}
\end{equation}
with the normalized state of $j$-th atom,
\begin{equation}
\begin{aligned}
|A'_j(t_1)\rangle
=
e^{i\hat{\xi}'_{j}(\delta t)}
e^{i\hat{\xi}_{j}}|A_j(t_1)\rangle\,.
\end{aligned}
\end{equation}
This presentation shows that atoms are identical and independent from each others. Thus, the single atom model in this paper is valid to describe the practical system of many atom being inside the interaction cavity simultaneously.

\section{conclusion}
We have shown that nanoparticle can be used as a light source to make Ramsey-Bord\'{e} atoms interferometer. The proposal is based on the cavity QED system which consists of a single atom and a single nanoparticle. The nanoparticle is illuminated by the external laser pulses, and then nearly-resonantly excite one of the cavity modes. This cavity mode generates  atomic ac Stark effect, and then the atomic KD scattering if the input laser is the pulsed one. The unitary transformations approach is used to derive such a KD scattering. It is proved that the present single atom model is also valid for the case that many atoms are simultaneously inside the interaction cavity, because the atomic internal state is not excited. Atomic KD scattering in the classical standing wave of lights is widely used to make the laser-pulsed atom interferometers. Thus, the present cavity system could be also feasible to generate atom interferometers. A slight difference is that the present atom interferometer is coupled with the centre-of-mass motion of nanoparticle, and thus could be developed as a tool to measure the classical or nonclassical motions of freely falling nanoparticle.
Once the motion of nanoparticle is exactly detectable, the freely falling nanoparticle (in the stage of $t_1-t_0$) may be further developed for the other uses~\cite{Romero-Isart-Science}, for example, studying the frequency-mixed effects of nanoparticle's nonlinear optics. In addition to nanoparticle detections, this study provides an alternative way to measure the classical or nonclassical motions of some other condensed particles, such as the macromolecule, which could be also used as the light source to make atom interferometers.

\textbf{Acknowledgments}:
This work was supported partly by the National Natural
Science Foundation of China, Grant No. 12047576.


\end{document}